\documentclass[aps,prb,twocolumn,showpacs,preprintnumbers,amsmath,amssymb, superscriptaddress, bibnotes]{revtex4}
\usepackage{graphicx}
\usepackage{dcolumn}
\usepackage{bm}
\usepackage{color}
\begin{document}

\title{Pressure phase diagram and quantum criticality of CePt$_2$In$_7$ single crystals}

\author{V. A. Sidorov}
\affiliation  {Los Alamos National Laboratory, Los Alamos, New Mexico 87545, USA}
\affiliation  {Institue for High Pressure Physics, Russian Academy of Sciences, 142190 Troitsk, Moscow, Russia}
\author{Xin Lu}
\email[Corresponding author: ]{xinluphy@gmail.com}
\affiliation  {Los Alamos National Laboratory, Los Alamos, New Mexico 87545, USA}
\affiliation  {Center for Correlated Matter, Zhejiang University, Hangzhou, 310058, China}
\author{T. Park}
\affiliation {Los Alamos National Laboratory, Los Alamos, New Mexico 87545, USA}
\affiliation {Department of Physics, Sungkyunkwan University, Suwon 440-746, Korea}
\author{Hanoh Lee}
\author{P. H. Tobash}
\author{R. E. Baumbach}
\author{F. Ronning}
\author{E. D. Bauer}
\author{J. D. Thompson}
\affiliation {Los Alamos National Laboratory, Los Alamos, New Mexico 87545, USA}

\begin{abstract}
We report the temperature-pressure (T-P) phase diagram of CePt$_2$In$_7$ single crystals, especially the pressure evolution of the antiferromagnetic order and the emergence of superconductivity, which have been studied by electrical resistivity and ac calorimetry under nearly hydrostatic environments. Compared with its polycrystalline counterpart, bulk superconductivity coexists with antiferromagnetism in a much narrower pressure region. The possible existence of textured superconductivity and local quantum criticality also are observed in CePt$_2$In$_7$, exhibiting a remarkable similarity with CeRhIn$_5$.
\end{abstract}

                            
\maketitle

The so called 115 heavy-Fermion family CeMIn$_5$ has attracted interest due to the intricate relationship between antiferromagnetism (AFM) and superconductivity (SC) that is found in them.\cite{ThompsonReview1,ThompsonReview2}  The increased superconducting transtion temperature T$_c$ ($\approx$ 2.1 K for CeRhIn$_5$ under pressure, 2.3 K for CeCoIn$_5$ at ambient pressure) compared with its cubic buliding block CeIn$_3$ (T$_c\approx$ 0.2 K under pressure) is believed to be due to the reduced dimensionality of CeIn$_3$ conducting layers that are separated by an MIn$_2$ layer in the 115 family. Though forming in a slightly different tetragonal structure, CePt$_2$In$_7$ is a cousin of CeMIn$_5$ in a broader family of compounds Ce$_m$M$_n$In$_{3m+2n}$, where m and n is the number of CeIn$_3$ and MIn$_2$ layers, respectively. From measurements on polycrystalline samples, CePt$_2$In$_7$ is antiferromagnetic with T$_N\approx$ 5.5 K at ambient pressure and superconductivity emerges under pressure with the suppression of AFM around a possible quantum critical point (QCP). \cite{EBauer127PRB10}  CeIn$_3$ layers in CePt$_2$In$_7$ are further distant from each other compared with those in CeMIn$_5$, giving a more two dimensional Fermi surface as revealed in quantum oscillation measurements.\cite{dHVa12PRB} However, its maximum T$_c\approx$ 2.1 K at 3.0 GPa is comparable to that in CeMIn$_5$, indicating that hybridization between f- and conduction- electrons has complicated the effect of reduced dimensionality on superconductivity.

CePt$_2$In$_7$ single crystals have similar physical properties as in the polycrystalline samples.\cite{PTobash12JPhys} However, NMR measurements on CePt$_2$In$_7$ have observed a distinction between single-crystal and polycrystalline samples. Coexistence of commensurate (CM) and incommensurate (ICM) AFM has been found in single crystals at ambient pressure and the CM-AFM is stabilized with pressure up to 2.4 GPa. \cite{HSakai11PRB} In contrast, only CM-AFM is observed in polycrystalline samples.\cite{NCurro10PRB} This difference has been interpreted to arise from the enhanced CM-AFM volume fraction caused by strain at the grain boundaries in the polycrystalline samples. It is thus desirable to revisit pressure measurements on CePt$_2$In$_7$ single crystals and to investigate its intrinsic pressure phase diagram. 

In this article, we report electrical resistivity and heat capacity measurements on CePt$_2$In$_7$ single crystals under nearly hydrostatic pressure environments. Superconductivity also emerges with the suppression of AFM order in single crystals, similar to the results reported before. However, bulk SC and AFM coexists in a narrower pressure range compared with the polycrystalline samples. The upper critical field and quantum critical behavior are also carefully studied.

CePt$_2$In$_7$ single crystals were grown from In flux as described elsewhere,\cite{PTobash12JPhys} with plate-like faces perpendicular to the c-axes, and they were screened by SQUID magnetometry to ensure the absence of free In. The crystals were mounted in two types of pressure cells: a toroidal cell with a glycerol-water mixture as the pressure medium for pressures up to 5.3 GPa and an indenter cell with silicon fluid or Daphne oil as the pressure medium for experiments up to 3.2 GPa. The nearly hydrostatic pressure inside the pressure cell at low temperatures was calibrated by the narrow superconducting transition temperature of a Pb manometer. The electrical resistivity was measured by the normal four-contact method with the current flowing in the ab plane. The heat capacity C$_{ac}$ of CePt$_2$In$_7$ under pressure was derived from ac calorimetry. In this technique, a small temperature oscillation $\Delta T$ generated by a heater glued to one face of the crystal is converted to an ac voltage signal by a chromel-AuFe (0.07\%) thermocouple fixed on the opposite side and C$_{ac}\propto \frac{1}{\Delta T}$. In order to calculate the specific heat of CePt$_2$In$_7$, a contribution from the pressure medium in the cell was subtracted from the total measured specific heat.\cite{Vladimir10ac} 
																																	
\begin{figure*}[http]
\includegraphics[angle=0,width=0.95\textwidth]{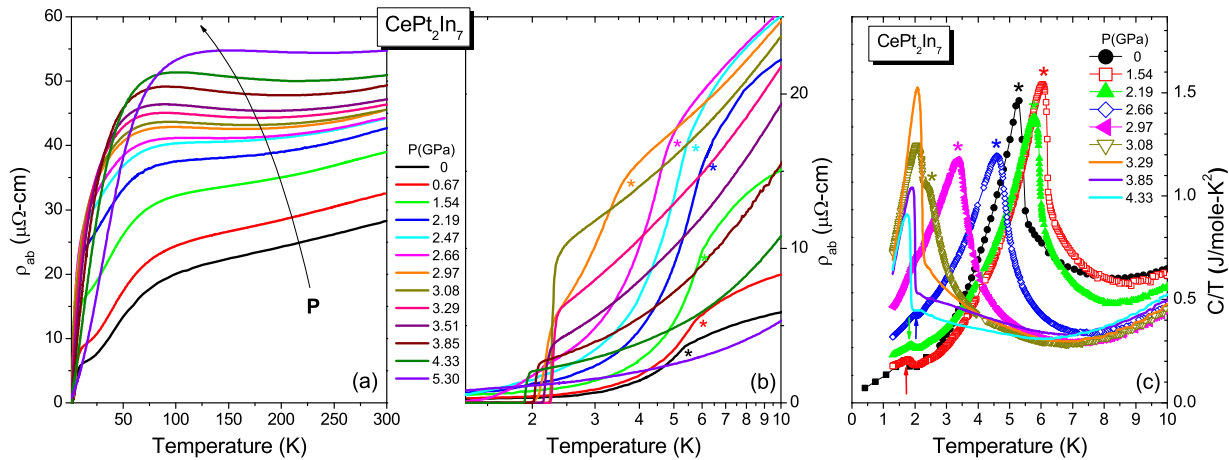}
\vspace{-12pt} \caption{\label{pressure-behavior} (Color online) Temperature-dependent (a) electrical resistivity $\rho(T)$  (the low temperature part is zoomed in for a clearer view in (b) with temperature on a log scale) and (c) specific heat $C/T$ of CePt$_2$In$_7$ single crystals as a function of pressure up to 4.33 GPa. The antiferromagnetic transition is marked by $\star$ in (b) and (c) and the specific heat jump due to the impurity phase CePt$_2$ is indicated by the arrows in (c).}
\end{figure*}

Figure \ref{pressure-behavior} (a) and (c) show the temperature-dependent resistivity and heat capacity of CePt$_2$In$_7$ single crystals as a function of pressure. At ambient pressure, both the sharp kink in the resistivity (or peak in $d\rho(T)/dT$) and the peak in specific heat at 5.3 K are consistent with its reported AFM N\'{e}el temperature T$_N$. We note that the specific heat curve $C_{ac}/T$ measured in the toroidal cell has a small jump around 1.7 K, indicating a minor CePt$_2$ impurity phase ($< 1\%$) in the sample as in [\onlinecite{PTobash12JPhys}], however, this impurity phase is absent in the resistively measured crystal and does not affect our major experimental results. With increased pressure on CePt$_2$In$_7$, its electrical resistivity at room temperature increases monotonically, while the rollover around 100 K develops into a broad local maximum clearly signaling the onset of coherence between Ce ions and a shoulder at 25 K gradually weakens and finally disappears. As shown in Fig. \ref{pressure-behavior} (c), the AFM transition initially moves to higher temperature, reaching its maximum at 1.7 GPa, and then is suppressed at higher pressures.

\begin{figure}
\includegraphics[angle=0,width=0.46\textwidth]{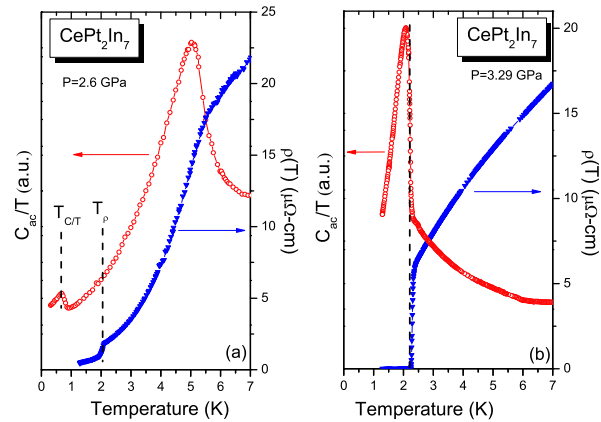}
\vspace{-12pt} \caption{\label{Figure2SC} (Color online) Temperature dependence of the electrical resistivity and heat capacity of CePt$_2$In$_7$ single crystals where the applied pressure is (a) 2.6 GPa and (b) 3.29 GPa, respectively. In (a), $T_{\rho}$ and T$_{C/T}$ determined by the resistivity drop and heat capacity jump show a significant difference while they are comparable  as indicated by the dashed line in (b). Note that the crystal measured at 2.6 GPa in the indenter cell has no CePt$_2$ impurity phase. }
\end{figure}

Both resistivity and ac calorimetry measurements provide clear evidence on the emergence of superconductivity with the suppression of AFM ordering as shown in Fig. \ref{pressure-behavior} (b) and (c). However, they display different behaviors in the pressure region where superconductivity and AFM coexist: If we compare the superconducting transitions at 2.6 GPa as shown in Fig.\ref{Figure2SC} (a), where AFM orders around 5 K, there is a sudden drop of resistivity at 2 K followed by a long tail indicating the onset of filamentary superconductivity. In contrast, a small peak around 0.6 K in the specific heat curve signals the establishment of bulk superconductivity. This is similar to results reported for CeRhIn$_5$ where textured filamentary superconductivity emerges before bulk superconductivity develops in the presence of a coexisting AFM order.\cite{TusonTexturedSC} Measurements of resistive transitions for current along different crystallographic directions are needed to confirm the presence of the textured SC, however, it is challenging due to the small size of our crystals. When AFM order is completely suppressed, the superconducting transition is sharp and both measurements mark a comparable transition temperature, as displayed in Fig. \ref{Figure2SC} (b) under the pressure of 3.29 GPa.

\begin{figure}
\includegraphics[angle=0,width=0.46\textwidth]{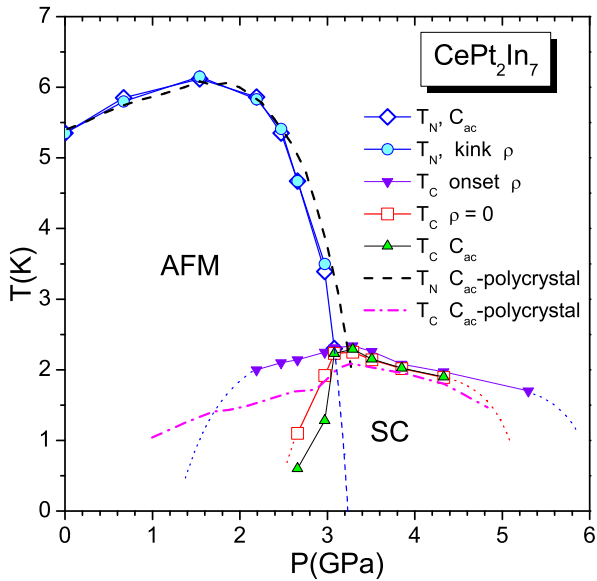}
\vspace{-12pt} \caption{\label{PhaseDiagram} (Color online) Temperature-pressure (T-P) phase digram of CePt$_2$In$_7$ single crystals in comparison with polycrystalline samples. The antiferromagnetic T$_N$ shows a similar pressure response but superconductivity determined by ac calorimetry measurements on single crystals (green triangles) occupies a much narrower region in the phase diagram relative to polycrystalline samples (dash-dotted line).}
\end{figure}

The temperature-pressure (T-P) phase diagram of CePt$_2$In$_7$ single crystals is summarized in Fig. \ref{PhaseDiagram} and compared to the polycrystalline samples. The antiferromagnetism evolves with pressure in almost the same way, but bulk superconductivity is present with AFM only in a much narrower pressure range from heat capacity measurements on single crystals. Notice that the dome of the onset of superconductivity in single crystals spans a similar pressure range as the bulk SC in polycrystalline CePt$_2$In$_7$ in Fig. \ref{PhaseDiagram}. One possible scenario to explain this distinction is that grain-boundary induced strain in the polycrystalline samples enhances the volume fraction of CM-AFM as discussed in [\onlinecite{HSakai11PRB}], which may favor conditions for the emergence of superconductivity compared with the ICM-AFM structure. It is much like the case of CeRhIn$_5$, where a new magnetic structure $Q_2=$(1/2, 1/2, 0.391) starts to appear with the original AFM wavevector Q$_1=$(1/2, 1/2, 0.326) right around the resistive superconducting onset temperature and completely replaces Q$_1$ below the bulk SC T$_c$.\cite{TusonTexturedSC} Further investigations of the effect of uniaxial pressure on CePt$_2$In$_7$ would be useful. The AFM N\'{e}el temperature is expected to reach zero around P$_C \sim$3.2 GPa from the extrapolation, approaching the assumed AFM quantum critical point (QCP). However, the possible QCP is embedded in the emerging superconducting dome and it is thus challenging to investigate its nature of the QCP.

\begin{figure}
\includegraphics[angle=0,width=0.46\textwidth]{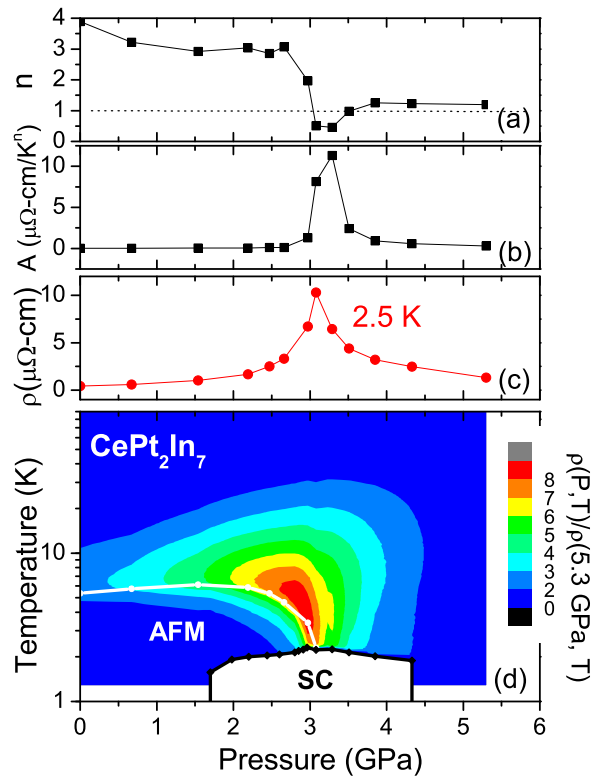}
\vspace{-12pt} \caption{\label{Figure4QCP} (Color online) Pressure dependence of the fitting parameters (a) n and (b) A for the electrical resistivity at low temperatures with the formula $\rho(T)=\rho_0+A T^n$. (c) The residual resistivity at 2.5 K (red dots), which peaks at 3 GPa. (d) The T-P contour color plot of isothermal resistivity $\rho(T,P)/\rho(T, 5.3 GPa)$, suggesting enhanced quasiparticle scattering around 3 GPa.}
\end{figure}

We first analyzed the low tempeature resistivity of CePt$_2$In$_7$ above its superconducting transition by fitting it to $\rho=\rho_0+AT^n$, where $\rho_0$, n, A are fitting parameters and the dependences of n and A on pressure are plotted in Fig.\ref{Figure4QCP} (a) and (b), respectively. Even though the fitting parameters are sensitive to the temperature range in consideration, a general trend is that they all exhibit an anomaly around P$_C$: the power-law exponent n is less than 1  and the $T^n$ coefficient A has a peak around 3.2 GPa as shown in Fig. \ref{Figure4QCP} (a) and (b). A similar sublinear temperature dependence of resistivity associated with a QCP also is found in CeRhIn$_5$ and its exact origin is still an open question.\cite{TusonRh115-2} Interestingly, we notice a negative residual resistivity $\rho_0$ at P$_C$ (not shown), which is obviously unphysical and likely indicates the presence of an additional minor energy scale at low temperatures. To avoid this negative $\rho_0$ issue, we simply take the resistivity value at 2.5 K as an estimate of the residual scattering rate and track its evolution with pressure as shown in Fig.\ref{Figure4QCP} (c); it also peaks around P$_C$ reflecting significantly enhanced quasiparticle scattering in CePt$_2$In$_7$ due to stronger spin fluctuations around the QCP. In Fig. \ref{Figure4QCP} (d), the isothermal resistivity as a function of pressure is normalized to the corresponding value at 5.3 GPa, $\rho(T,P)/\rho(T, 5.3 GPa)$, and mapped over a broad temperature range. The maximum in this ratio also centers around the pressure where T$_c$ is a maximum, which suggests an intimate relationship between enhanced spin fluctuations at the QCP and unconventional superconductivity in CePt$_2$In$_7$. Fig. \ref{Figure4QCP} (d) also demonstrates the extent of strong scattering in a wide range in the T-P diagram, which has been interpreted in CeRhIn$_5$ as a form of local quantum criticality.\cite{TusonRh115-2}  

\begin{figure}
\includegraphics[angle=0,width=0.46\textwidth]{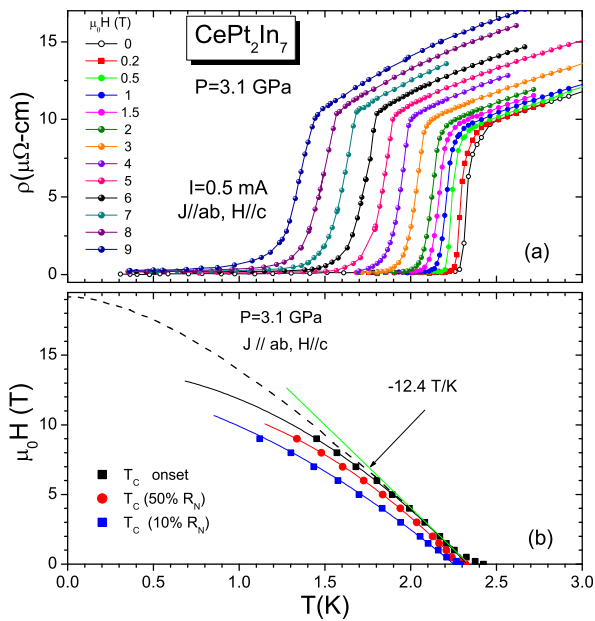}
\vspace{-12pt} \caption{\label{Figure5Hc2} (Color online) (a) Temperature dependence of the electrical resistance for CePt$_2$In$_7$ under magnetic fields up to 9 T where the pressure is 3.1 GPa and the field is applied in c-axis direction with current in the ab plane. (b) The upper critical field Hc$_{2}$ of CePt$_2$In$_7$ single crystals at 3.1 GPa determined by the superconducting resistance transition. Hc$_{2}$ estimated from the WHH formula is plotted as a dashed line.}
\end{figure}

Figure \ref{Figure5Hc2} (a) plots the temperature dependent resistivity of a CePt$_2$In$_7$ single crystal at 3.1 GPa and in a magnetic field applied perpendiclar to the ab plane. The superconducting transition is only reduced to 1 K at 9 T from the initial zero-field T$_c\sim$ 2.3 K, which suggests a large upper critical field Hc$_2$ due to an enhanced electron mass $m^*$ and reduced coherence length $\xi$ typical of heavy-fermion superconductors. Three different criteria are adopted to determine the superconducting T$_c$ quantitatively as a function of field: the onset temperature of the resistance drop and the temperatures where the resistance reaches 50\% and 10\% of of its normal state value. They are plotted in Fig. \ref{Figure5Hc2} (b). The initial slope $-dHc_{2}/dT_c\sim$ -12.4 T/K is close to -15 T/K observed by Muramatsu et al. for CeRhIn$_5$ with the same orientation of the magnetic field near its pressure-tuned QCP.\cite{CeRhIn5Hc2} The extrapolated Hc$_2$(0) $\sim$ 15 Tesla is lower than that ($\sim$ 19 Tesla) estimated from the Werthamer-Helfand-Hohenberg (WHH) formula for orbital pair-breaking in the dirty-limit. This difference suggests that the upper critical field may be limited by Pauli paramagnetic pair breaking, which is also suggested for CeRhIn$_5$.\cite{CeRhIn5ParkNJPhys} According to the equations $-dHc_{2}/dT_c\propto m^{*2} T_c$ and $\xi=\sqrt{\Phi_0/{2\pi Hc_{2}}}$, it yields $m^*=$ 160 m$_e$ and $\xi=$ 46.8 \AA. Intestingly, a sub-linear temperature dependence of the resistance is robust against fields to at least 9T and the magnetoresistance is positive. These raise the possibility of a field-induced T=0 magnetic transition inside the superconducting dome.

In summary, we have constructed the pressure-temperature phase diagram of CePt$_2$In$_7$ single crystals from both electrical resistivity and ac calorimetry measurements. These experiments find a narrower pressure range of bulk superconductivity coexsiting with AFM compared with results on polycrystalline samples. Further studies are needed to probe the order parameter symmetry of the pressure-induced superconductivity and to search for possible hidden magnetism between 3.0 GPa and 3.2 GPa as in CeRhIn$_5$.\cite{TusonRh115-1}

We are grateful for valuable discussions with J. X. Zhu. Work at Los Alamos was performed under the aupices of the U. S. Department of Energy, Division of Materials Science and Engineering and suported in part by the Los Alamos LDRD program. V.A.S. acknowledges a support from Russian Foundation for Basic Research (RFBR Grant No. 12-02-00376) and Program of the Physics Depertment of RAS on Strongly Correlated Systems. T.P. acknowledges a support from NRF grant (No. 2010-0026762 and 2010-0029136).

\bibliographystyle {apsrev}

\end{document}